\documentclass[aps,prl,reprint,showpacs,superscriptaddress]{revtex4-1} 
\usepackage{graphicx}  
\usepackage{dcolumn}  
\usepackage{bm}      
\usepackage{amssymb}   
\usepackage{bm}     
\usepackage{hyperref} 
\usepackage{dsfont}   
\usepackage{color}

\begin{document}

\title{Second-order temporal interference with thermal light: \\Interference beyond the coherence time}


\author{Yong Sup Ihn}
\email{yong0862@gmail.com}
\affiliation{Department of Physics, Pohang University of Science and Technology (POSTECH), Pohang, 37673, Korea}

\author{Yosep Kim}
\affiliation{Department of Physics, Pohang University of Science and Technology (POSTECH), Pohang, 37673, Korea}

\author{Vincenzo Tamma}
\email{vincenzo.tamma@port.ac.uk}
\affiliation{Faculty of Science, SEES and  Institute of Cosmology \& Gravitation, University of Portsmouth, Portsmouth, PO1 3QL, UK}

\author{Yoon-Ho Kim}
\email{yoonho72@gmail.com}
\affiliation{Department of Physics, Pohang University of Science and Technology (POSTECH), Pohang, 37673, Korea}
\date{\today}

\begin{abstract}
We report observation of a counter-intuitive phenomenon in multi-path correlation interferometry with thermal light. The intensity correlation between the outputs of two unbalanced Mach-Zehnder interferometers (UMZI) with two classically correlated beams of thermal light at the input  exhibits genuine second-order interference with the visibility of $1/3$. Surprisingly, the second-order interference does not degrade at all no matter how much the  path length difference in each UMZI is increased beyond the coherence length of the thermal light. Moreover, the second-order interference is dependent on the difference of the UMZI phases. These results differ substantially from those of the entangled-photon Franson interferometer which exhibits  two-photon interference dependent on the sum of the UMZI phases and the interference vanishes as the path  length difference in each UMZI exceeds the coherence length of the pump laser.  Our work offers deeper insight into the interplay between interference and coherence in multi-photon interferometry.
\end{abstract}



\maketitle


Two-photon interference or  second-order interference, in which interference is observed only in the correlation between two detectors,  has long been at the heart of quantum optics and it has its root in the Hanbury-Brown--Twiss (HBT) experiment \cite{HBT56a,Glauber63a}. The quintessential effect of the HBT experiment with thermal light is that the joint detection probability of the two detectors is twice as large when the two detectors ``click'' simultaneously  than that of the case when the two detectors ``click'' with a relative time delay bigger than the coherence time of the thermal light \cite{Morgan66,Arecchi66}. While the HBT effect with thermal light can be explained  as correlation of intensity fluctuations, quantum mechanically, it is understood as constructive interference between two indistinguishable two-photon detection probability amplitudes \cite{Fano61}.  HBT interferometry in recent years has become essential for a variety of studies in quantum physics, e.g., bunching and anti-bunching of photons, electrons, and atoms \cite{Grangier86,Henny99,Jeltes07}.

\begin{figure*}[t]
\includegraphics[width=6.8in]{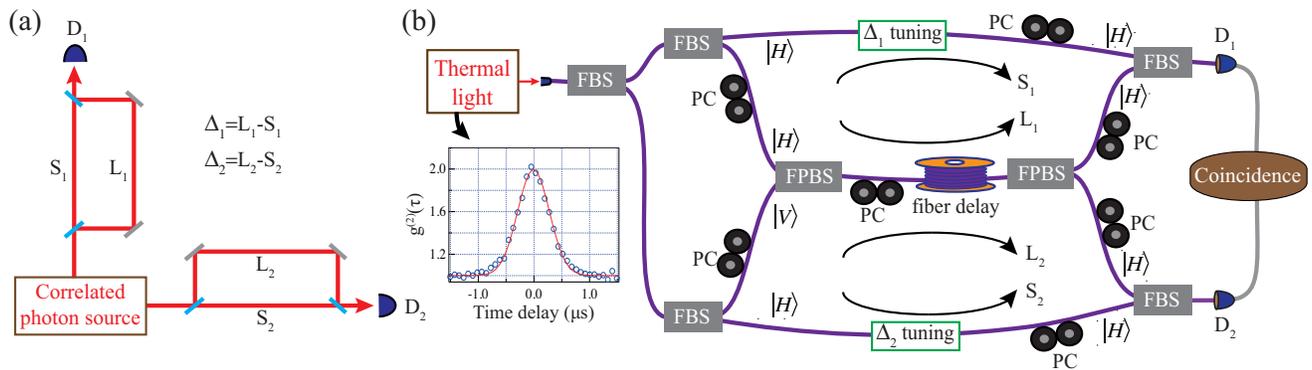}
\caption{(a) Proposed setup measuring correlation between the outputs of the two unbalanced Mach-Zehnder interferometers (UMZI). Similarly to the Franson interferometer, UMZI has the path length difference much larger than the coherence length of the input light so that no first-order interference is observed at the detectors $D_1$ and $D_2$. Unlike the Franson interferometer, we consider two classically correlated beams of light, produced by beam splitting of a thermal light beam.
(b) Schematic of the experimental setup. The thermal light beam is first split into a pair of correlated beams with a fiber beam splitter (FBS). The inset displays the measured $g^{(2)}(\tau)$ function of the thermal light having the full width at half-maximum coherence time $\tau_c=572$ ns. The UMZI is constructed by using fiber-optic delay lines. To mitigate the effect of random phase fluctuation between the two UMZI, the long paths $L_1$ and $L_2$ share the same fiber spool of 200 m, 400 m, 600 m, or 800 m. Note that a 120 m optical fiber delay is sufficient to completely remove the first order interference. The $L_1$ and $L_2$ paths are defined by the polarization states $|H\rangle$ and $|V\rangle$, respectively, by using a set of fiber polarizing beam splitters (FPBS). The delays $\Delta_1$ and $\Delta_2$ are tuned by piezoelectric actuators. PC refers to the fiber polarization controller. 
}\label{fig:setup}
\end{figure*}

In this Letter, we demonstrate experimentally a novel second-order temporal interference effect proposed theoretically in Ref.~\cite{Tamma16a}. Differently from the usual HBT effect,  we  show the emergence of sinusoidal second-order interference fringes which seems to contradict the common understanding of temporal coherence \cite{Shih86,Hong87,Baek07,Kwon09,Ra13a}. The essential idea of the experiment is shown in Fig.~\ref{fig:setup}(a). A pair of classically correlated beams is generated by beam splitting of a thermal light beam \cite{Cho10a,Cho10b}. Each beam is then sent through an unbalanced Mach-Zehnder interferometer (UMZI) with the path length difference between the long and short paths larger than the longitudinal coherence length $c\tau_c$ of the thermal light. The second-order correlation $ G^{(2)}(t_{1},t_{2})$ between the detectors $D_1$ and $D_2$ placed at the output of  UMZI is measured with a coincidence time window smaller than the coherence time $\tau_c$  of the thermal light, i.e. $|t_2-t_1| \ll \tau_c$.  The two UMZI satisfy the following conditions. First, the path length differences $\Delta_1=L_1-S_1$ and $\Delta_2=L_2-S_2$ are larger than the coherence length of the thermal light, i.e., $\Delta_{1},  \Delta_{2} \gg c\tau_{c}$. This condition ensures that there is no fist-order interference at the detectors $D_1$ and $D_2$. Second, the two UMZI are similar to each other in that the differences of the corresponding optical paths are small compared to the coherence length of the thermal light, i.e., $|L_{1}-L_{2}|,  |S_{1}-S_{2}|  \ll  c\tau_{c}$. Under these conditions, the correlation measurement picks up second-order interference due to the relative phase difference between the long $(L_1,L_2)$ path and the short $(S_1,S_2)$ path.  The  second-order correlation function $G^{(2)}(t_{1},t_{2})$  which is manifested in the  coincidence count rate is then given by \cite{Tamma16a}
\begin{equation}
G^{(2)}(|t_2-t_1| \ll \tau_c) \propto 3+\cos\left\{\frac{\omega}{c} (\Delta_{1}-\Delta_{2})\right\}.
\end{equation}

We now briefly compare the above results to those of the Franson interferometer in which the input photon pair is  energy-time entangled so that the interferometer serves as an apparatus to measure energy-time entanglement \cite{Franson89,Shih93a,Kwiat93a,Strekalov96}.  In our scheme, we consider two classically correlated beams of light, produced by beam-splitting of a thermal light beam. Nonetheless, temporal correlation between the long paths $(L_1,L_2)$  and the short paths $(S_1,S_2)$  do exist when the correlation measurement is performed at the coincidence time window  smaller than  the coherence time $\tau_c$ of the thermal light  \cite{Tamma16a}. 
Indeed, the second-order temporal interference phenomenon reported here emerges  from interference of two  effective probability amplitudes associated with two pairs of correlated paths $(L_1,L_2)$ and $(S_1,S_2)$.  Interestingly, the second-order interference does not degrade at all no matter how much the  path length difference in each UMZI ($\Delta_1$ and $\Delta_2$) is increased beyond the longitudinal coherence length $c\tau_c$ of the thermal light. This represents a counter-intuitive manifestation of second-order temporal coherence. Indeed, this is in stark contrast to  entangled-photon Franson interferometry in which the second-order interference is limited to the coherence length of the pump laser generating the energy-time entangled photons \cite{Shih93a,Kwiat93a,Strekalov96}.




The  experimental setup is schematically shown in Fig.~\ref{fig:setup}(b). First, the thermal light beam is generated by focusing a laser beam onto a rotating ground disk \cite{Cho10a,Cho10b}. An external cavity diode laser operating at 780 nm is used and it is frequency locked to the  $5S_{1/2}(F=3) - 5P_{3/2}(F'=4)$ transition of the $^{85}$Rb atomic energy levels. The rotating ground disk transforms the input coherent state with the Poissonian photon number statistics into the output Bose-Einstein photon number statistics having the coherence time of $\tau_c=572$ ns, corresponding to the coherence length of approximately 120 m in an optical fiber. The measured second-order correlation function  $g^{(2)}(\tau)$ is shown in the inset of Fig.~\ref{fig:setup}(b), demonstrating the photon bunching property  of the thermal light source. A thermal light source with a rather large coherence time of $\tau_c=572$ ns is used in the experiment to ensure that the coherence time well exceeds the timing resolution of the single-photon detectors (Perkin-Elmer SPCM-AQR) and the coincidence electronics. As the proposed second-order temporal interference originates from the photon bunching shown in the $g^{(2)}(\tau)$, it is essential that the combined timing resolution of the detectors and electronics does not degrade the bunching effect.   

The horizontally polarized thermal light beam is first split into a pair of correlated beams with a fiber beam splitter (FBS). Each beam is then sent through an unbalanced Mach-Zehnder interferometer (UMZI) with a long and a short optical fiber path. The UMZI consists of FBS, fiber polarizing beam splitters (FPBS), fiber polarization controllers (PC), and optical fibers. The short paths, $S_1$ and $S_2$, each contain a 1 m long optical fiber and  a free-space delay line, labeled as $\Delta_1$ tuning or $\Delta_2$ tuning, controlled by a piezo actuator (Thorlabs AE0505D16F) for phase modulation. The long paths, $L_1$ and $L_2$, each include a long fiber spool of length 200 m, 400 m, 600 m or 800 m. Note that, since the coherence time of the thermal light is 572 ns which corresponds to the coherence length of 120 m in an optical fiber, a 200 m long optical fiber spool is more than sufficient to completely remove any first-order interference at the output of the UMZI. To mitigate the effect of random phase fluctuation between the two UMZI, mostly arising from the long optical fiber spools, the long paths $L_1$ and $L_2$ of the two UMZI  physically share the same fiber spool. The $L_1$ and $L_2$ paths, instead, are defined by the polarization states $|H\rangle$ and $|V\rangle$, respectively, by using PC and FPBS. Finally, the $\Delta_{1}$ and $\Delta_{2}$ delays are scanned by applying voltages to the piezo actuators while observing the single and coincidence counting rates of the two detectors $D_1$ and $D_2$. The coincidence time window in the experiment is set at 15 ns.

\begin{figure}[t]
\includegraphics[width=3in]{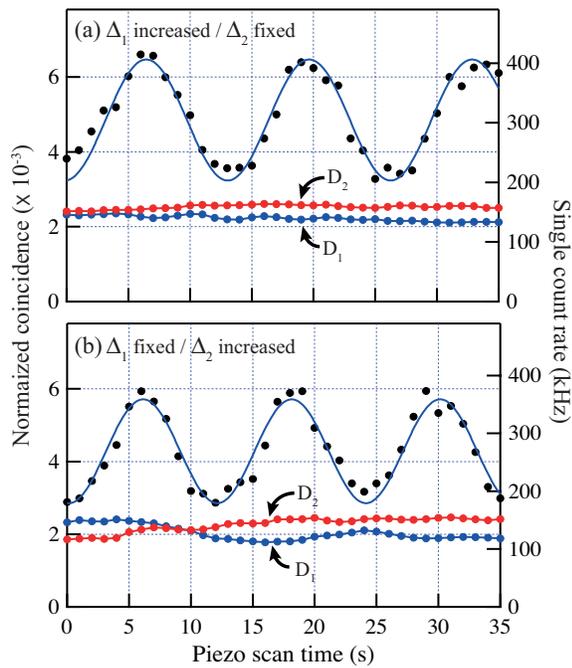}
\caption{Second-order interference of thermal light with the fiber delay line of 200 m, see Fig.~\ref{fig:setup}(b). The piezo is continuously scanned linearly at 63 nm/s for increasing/decreasing $\Delta_1$ and $\Delta_2$. Each data point is accumulated for 1 s. It is clear that there is no first-order interference, as evidenced in the $D_1$ and $D_2$ count rates. The normalized coincidence however exhibits second-order interference. The solid lines are sinusoidal fits to the data with the visibility fixed at the theoretical maximum value of 1/3.}\label{fig:single}
\end{figure}

We now report the experimental observation of the second-order temporal interference with thermal light. First, a 200 m long fiber spool is used for the long paths $L_1$ and $L_2$ and the second-order interference is observed by scanning one of the piezo actuators in the short paths while the other is fixed. The piezo is scanned by applying a triangular voltage pattern up to 30 V at the rate 0.6 V/s, corresponding to increasing or decreasing of the delay $\Delta_1$ or $\Delta_2$ at the rate of 63 nm/s. The corresponding experimental data are shown in Fig.~\ref{fig:single}. To accurately show the genuine second-order interference effect, the coincidence count rate $N_c$ is normalized, i.e., $N_c/\sqrt{N_1 N_2}$, where $N_1$ ($N_2$) is the single count rate of $D_1$ ($D_2$). It is evident from the data in Fig.~\ref{fig:single} that there is no first-order interference while second-order interference is present. Moreover the second-order interference data fit nicely to the predicted sinusoidal fringe with the visibility of 1/3 in Eq.~(1).  

One of the interesting outcomes  of Eq.~(1) is that  the second-order temporal interference of thermal light is dependent on the difference of the UMZI phases, unlike the entangled photon case in which the interference is dependent on the sum of the UMZI phases. To test this scenario, we now scan both $\Delta_1$ and $\Delta_2$ simultaneously either in the opposite directions or in the same directions. According to Eq.~(1), when the phase difference $\Delta_1-\Delta_2$ is scanned, the interference fringe will occur twice as fast as the case with increasing/ decreasing of $\Delta_1$ or $\Delta_2$. Therefore, the piezo actuators for changing $\Delta_1$ and $\Delta_2$ are now scanned at the half speed as before at 0.3 V/s, corresponding to  32 nm/s. 

\begin{figure}[t]
\includegraphics[width=3in]{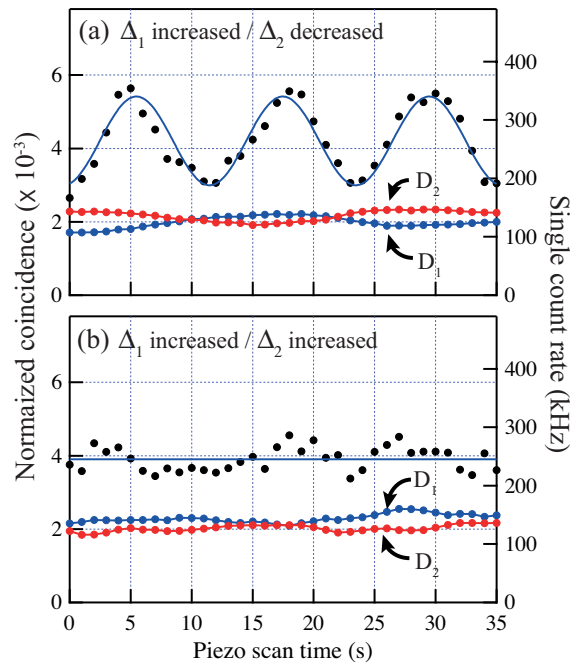}
\caption{Second-order interference of thermal light with the fiber delay line of 200 m when both $\Delta_1$ and $\Delta_2$ are simultaneously scanned, see Fig.~\ref{fig:setup}(b). The piezo scanning speed is now reduced to 32 nm/s. Each data point is accumulated for 1 s. 
(a) When $\Delta_1$ and $\Delta_2$ are scanned in the opposite direction, second-order interference is observed. The solid line is a sinusoidal fit to the data with the visibility fixed at the theoretical maximum value of 1/3. The data show that the second-order interference period in this condition is reduced by half compared to the cases of Fig.~\ref{fig:single}.
(b) There is no second-order interference when $\Delta_1$ and $\Delta_2$ are scanned in the same direction. Note that the second-order interference behavior reported here is quite the contrary to the case of entangled-photon Franson interferometry in which the $\lambda/2$ effect is observed when $\Delta_1$ and $\Delta_2$ are scanned in the same direction.}\label{fig:both}
\end{figure}

The experimental data for this case are shown in Fig.~\ref{fig:both}. When the phase difference $\Delta_1-\Delta_2$ is scanned by increasing $\Delta_1$ and decreasing $\Delta_2$ simultaneously at the same speed, the expected second-order interference with the visibility of 1/3 is clearly observed, see Fig.~\ref{fig:both}(a). However, when the phase sum $\Delta_1+\Delta_2$ is scanned by increasing $\Delta_1$ and $\Delta_2$ simultaneously at the same speed, there is no second-order interference, Fig.~\ref{fig:both}(b). In all cases, the single photon detection rates are constant, demonstrating the genuine second-order nature of the observed interference. Note again that the case reported here is in stark contrast to the case of entangled-photon Franson interference, in which high-visibility two-photon interference occurs at  half the wavelength of the photons when the UMZI phase sum is scanned.

\begin{figure}[t]
\includegraphics[width=3in]{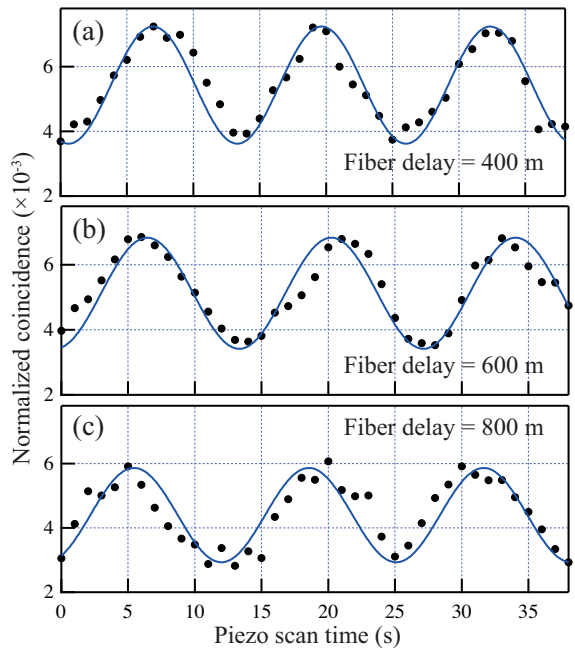}
\caption{Second-order interference of thermal light with the fiber delay line of 400 m, 600 m, or 800 m, see Fig.~\ref{fig:setup}(b). The piezo is continuously scanned linearly at 63 nm/s for increasing $\Delta_2$ and $\Delta_1$ is fixed. Each data point is accumulated for 1 s. The solid lines are sinusoidal fits to the data with the visibility fixed at the theoretical maximum value of 1/3. No reduction in visibility is observed. This is in stark contrast to the entangled-photon Franson interferometry in which the second-order interference is limited to the pump coherence length.} \label{fig:long}
\end{figure}

Finally, another interesting fact we find from Eq.~(1) is that the second-order temporal interference surprisingly does not degrade at all with the increase of the path length difference in each UMZI, i.e., the interference is completely independent of the coherence time of thermal light. In other words, the second-order temporal interference may be observed even at an extremely large path length difference, orders of magnitude bigger  than the coherence length of the thermal light, between the long and short paths of UMZI. Again, this is in stark contrast to the entangled-photon Franson interferometry in which the second-order interference is limited to the pump coherence length.  To test this scenario, the 200 m long fiber spool used for the long paths of the UMZI is now replaced with a longer fiber spool of 400 m, 600 m, or 800 m. Also, $\Delta_2$ is scanned at 63 nm/s while $\Delta_1$ is fixed. The experimental data are shown in Fig.~\ref{fig:long}. The normalized coincidence data show the expected second-order interference with the same visibility of 1/3. The slight reduction of the normalized coincidence rate for a longer fiber is due to the absorption loss at the optical fiber.

 
The  second-order temporal interference of thermal light reported here is a consequence of contributions from all the possible joint-detection amplitudes  associated with any possible pair of paths of the thermal light field components to the two detectors, see Fig.~\ref{fig:setup}(a) \cite{Tamma16a}.  However, only the amplitudes overlapping in time within the coherence time of the thermal light can interfere and, due to the chaotic nature of continuous-wave thermal light, all possible pairs of the thermal light components leading to the coincidence detection contribute to the interference.  Remarkably, the sum of all these contributions leads to interference between the effective probability amplitudes associated with two pairs of paths  $(L_1,L_2)$ and $(S_1,S_2)$ independently of how much the corresponding time delays, $\Delta_1/c$ and $\Delta_2/c$, are increased beyond the coherence time of the thermal light. These two effective probability amplitudes, interestingly, depend on the difference between the phase delays in the long and the short paths, respectively. The interference between these two effective amplitudes therefore leads to the sinusoidal oscillation in terms  of the relative path difference $\Delta_1 - \Delta_2$ with the visibility of 1/3 as shown in Eq.~(1).

We point out that these results are  fundamentally of different origin from those of the entangled-photon Franson interferometer, where the second-order interference is the result of interference between probability amplitudes associated with a single pair of entangled photons taking either the $(L_1,L_2)$ or the $(S_1,S_2)$ path. Indeed, the Franson interferometer  exhibits  two-photon interference dependent on the sum of the UMZI phases, $\Delta_1 + \Delta_2$, and the interference vanishes as the path length difference in each UMZI exceeds the coherence length of the pump laser \cite{Franson89,Shih93a,Kwiat93a,Strekalov96}. It is also worth pointing out that it is possible to achieve 100\% visibility by measuring correlation of the photon number fluctuations instead of measuring correlation of the intensities \cite{Scarcelli04a,Ferri05b,Agafonov09,Chen13}. The correlation of the photon number fluctuations at the output of the UMZI  is  calculated to be $\langle\Delta n_{1}\Delta n_{2}\rangle \propto 1+\cos\{\frac{\omega}{c} (\Delta_{1}-\Delta_{2})\}$ \cite{Tamma16a}.

In summary, we have reported observation of second-order temporal interference with thermal light. The intensity correlation between the outputs of two  UMZI is shown to exhibit second-order interference with the visibility of $1/3$ for the thermal light. The interference is shown to be dependent on the UMZI phase difference, unlike the Franson interferometer which exhibits the dependence on the UMZI phase sum. Furthermore, the second-order interference does not degrade at all no matter how much the   path length difference is increased beyond the coherence length of the thermal light. This is due to the fact that  photon bunching   of continuous-wave thermal light provides  second-order coherence between the $(L_1,L_2)$ path and the $(S_1,S_2)$ path,  regardless of the path length differences $\Delta_1$ and $\Delta_2$. The phenomenon demonstrated here, for instance, can be used to measure an unknown longitudinal phase difference between two remote locations, analogous to recent demonstrations of spatial second-order interference with two remote double slits \cite{Cassano17,Peng16,Dangelo17}. We thus believe that our work offers deeper insight into the interplay between interference and coherence in multi-photon interferometry and provides potential applications of  interferometry with classical light in metrology and imaging \cite{Ra13a,Ra13b,Tamma14b,Tamma15a}.



This work was supported by the National Research Foundation of Korea (Grant Nos. 2016R1A2A1A05005202 and 2016R1A4A1008978) and the KIST Open Research Program. YK  acknowledges the Global Ph.D. Fellowship by the National Research Foundation of Korea (Grant No. 2015H1A2A1033028). VT was partially supported by the Army Research Laboratory (W911NF-17-2-0179).

\end{document}